%% file: main.tex
\documentclass[conference]{IEEEtran}
\IEEEoverridecommandlockouts
\usepackage{cite}
\usepackage{amsmath,amssymb,amsfonts}
\usepackage{algorithmic}
\usepackage{graphicx}
\usepackage{textcomp}
\usepackage{xcolor}
\usepackage{fontawesome}
\usepackage{pgfplots}
\pgfplotsset{compat=1.18}
\usepackage[para]{footmisc}
\usepackage{dblfloatfix} 
\usepackage{tcolorbox}
\usepackage{multirow}
\usepackage{float}
\usepackage[normalem]{ulem}
\def\BibTeX{{\rm B\kern-.05em{\sc i\kern-.025em b}\kern-.08em
    T\kern-.1667em\lower.7ex\hbox{E}\kern-.125emX}}

 \newenvironment{quote1}{%
   \list{}{%
     \leftmargin%0.5cm   % this is the adjusting screw
        \rightmargin%0.5cm
   }
   \item\relax
}
{\endlist}

 \newenvironment{quote2}{%
   \list{}{%
     \leftmargin\parindent%0.5cm   % this is the adjusting screw
        \rightmargin\parindent%0.5cm
   }
   \item\relax
}
{\endlist}

\newcommand{\myquote}[2]{\begin{quote1}{\small\faComment}~\emph{#1}\end{quote1}} 
\newcommand{\rqquote}[1]{\begin{quote2}{#1}\end{quote2}} 

\begin{document}

\title{The Other Side of the Screen: Motivations to Watch and Engage in Software Development Live Streams}

\author{\IEEEauthorblockN{1\textsuperscript{st} Ella Kokinda}
\IEEEauthorblockA{\textit{School of Computing} \\
\textit{Clemson University}\\
Clemson, SC, USA \\
ekokind@gmail.com}
\and
\IEEEauthorblockN{2\textsuperscript{nd} D. Matthew Boyer}
\IEEEauthorblockA{\textit{Dept. of Engineering and Science Education} \\
\textit{Clemson University}\\
Clemson, SC, USA \\
dmboyer@clemson.edu}
}

\maketitle

\begin{abstract}
    \textbf{Background}:  With the popularity of live streaming platforms at an all-time high, and many people turning to alternative venues for educational needs, this full research paper explores the viewership habits of software and game development live streams through the lens of informal education opportunities. \\

    \noindent\textbf{Purpose}: We investigate why developers watch software and game development live streams to understand the educational and social benefits they derive from this emerging form of informal learning.\\

    \noindent\textbf{Methods}: We implement a mixed-methods study combining survey data from 39 viewers and nine semi-structured interviews to analyze motivations, perceptions, and outcomes of watching development live streams. \\

    \noindent\textbf{Findings}: This research finds that viewers are motivated by both educational and social factors, with community engagement and informal mentorship as key motivations. Additionally, we find that technical learning draws initial interest, but social connections and co-working aspects sustain long-term engagement. \\

    \noindent\textbf{Implications}: Live streaming serves as a valuable informal learning tool that combines self-directed technical education with community support, which suggests that developers can leverage these platforms for continuous learning and professional growth outside of or in addition to traditional educational structures.\\

    \noindent\textbf{Keywords}: Informal education, self-regulated learning, continuing education, community of practice
\end{abstract}

\begin{IEEEkeywords}
live streaming, informal learning, computer science education, community of practice, informal education, self education
\end{IEEEkeywords}

\input{sections/01_intro}
\input{sections/02_background}
\input{sections/03_method}
\input{sections/04_results}
\input{sections/05_discussion}

\input{sections/06_conclusion}

%%
%% Print the bibliography
%%
%\printbibliography
\bibliographystyle{IEEEtran}
\bibliography{biblo}

\appendix
\input{sections/07_surveyqs}
%Survey questions available via an online appendix: 

\end{document}

%% file: sections/01_intro.tex
\section{Introduction}
Access to streaming platforms has begun to change how people approach online education by offering access to real-time learning experiences in many domains - science, software and development, art, and music. Software and game development live streams emerged and formed into micro-communities where streamers openly share development processes and opinions, seek support and encouragement, and build a community of like-minded individuals \cite{kokinda2023streaming}. As these platforms become increasingly popular, we want to understand what motivates and draws viewers to participate in these types of streams and whether these streams have any perceived impact on viewers' development practices.  

Prior work in software and game development streams investigates from the streamer's perspective and the resulting lower barriers to entry for self-education, accountability for work, and community advocacy \cite{chen2021towards, faas2018watch, kokinda2023streaming}. For viewership, much of the prior work focuses on viewers of gaming streams, but still shows that stream viewers seek to gain knowledge \cite{sjoblom2017people, fraser2019sharing}. Additionally, in the context of video games, streamers and viewers take on the roles of mentors and apprentices and form learning communities, and the skill levels of the streamer can be influenced by the skill level of the viewer members \cite{burroughs2015esports,wulf2020watching, walker2014watching, greenberg2016interaction}. Understanding viewership through the lens of informal learning opportunities in software and game development streams will be important as live streaming platforms and alternative education venues continue to grow \cite{hiltz2005education, johnson2019impacts}.

This work aims to better understand viewers' motivations for watching and engaging with software and game development live streams, and where viewers find perceived benefit in watching and/or participating in live streams. Additionally, this work aims to build on prior work to create a more holistic view of the benefits and challenges of live streamed software development. Therefore, we ask the following research questions: 

\rqquote{{$RQ_1$:} What are the motivations of viewers of software and game development live streams?}
\rqquote{{$RQ_2$:} What are viewers' perceptions of software and game development streamers and how do they perceive their engagement with streamers?}
\rqquote{{$RQ_3$:} What are viewers' perceived beneficial outcomes of watching software and game development live streaming?}

The rationale behind each of these research questions is to understand the motivations to watch live streamed development content, how viewers perceive interactions in live streams, and how viewership positively impacts the individual choosing to watch this type of content. This information helps us understand if and how participating in software live streams, through passive viewership or active engagement in the streams, can act as a potential informal learning opportunity for viewers. 

%% file: sections/02_background.tex
\section{Background and Related Work}
This section discusses the background and related work on live streaming software development, live streaming viewership, and participation in online communities.

\subsection{Live Streaming Software Development}
While Twitch and YouTube Live are primarily known for gaming and entertainment content, they also host a growing and diverse array of creative and educational streams. These streams include artists, makers, musicians, and STEM-related broadcasts – including a category specifically for software and game development. Recent research indicates that developers who stream their work find it beneficial for self-education, accountability, and perceived skill enhancement \cite{kokinda2023streaming}. The community aspects of these streams are also significant, with viewers contributing both socially and technically to the content creators they engage with \cite{kokinda2023streaming}. Compared to prerecorded video content, live streaming offers a lower barrier for entry and a real-time learning experience that allows viewers to follow along with experts and knowledgeable individuals as they work \cite{chen2021towards}.

Research suggests that software development live streams could be effectively integrated into mainstream educational environments \cite{faas2018watch}. Live streaming offers distinct advantages over traditional prerecorded educational content, notably requiring less upfront preparation and investment \cite{chen2021towards}. This accessibility enables knowledge sharing from practitioners who may lack formal teaching credentials but possess valuable expertise. Additionally, streaming platforms provide an avenue for developer advocacy and insight into software development careers \cite{chattopadhyay2021developers}. The benefits of live streaming as an educational tool have been demonstrated across multiple fields. In medical education, for instance, live-streamed surgical procedures provide students with active engagement and immersive learning experiences that boost confidence through real-time observation and interaction  \cite{abu2022use, gandsas2023immersive}. Similar benefits appear in gaming communities, where live streams foster community participation, spark new interests, and create informal learning opportunities. These platforms have emerged as versatile tools for education across multiple disciplines.

For learners, live streaming provides unique benefits, particularly for those who prefer observational learning styles \cite{chen2021towards}. The format allows viewers to witness real-time problem-solving approaches and personalize their learning experience \cite{selwyn2007infromal}. However, current platform limitations, such as basic categorization systems, can make it challenging for viewers to find specific educational content or teaching styles. Despite these constraints, platforms continue to attract information-seeking viewers across various domains. Research indicates that learning outcomes remain consistent whether the content is delivered by novice or expert streamers, with streamer personality and approachability playing key roles in viewer engagement and community formation.

\subsection{Participation in Online Communities}
Participation in online comminutes has continued to be important in today's ever-increasing digital landscape, regardless of interests or professions. Online communities, particularly those that form on social networking sites and creative or open-source communities, offer a variety of benefits that span social, psychological, and professional domains \cite{malinen2015understanding}. These communities unite individuals to connect, share interests and experiences, and provide a space to access knowledge and well-being that may not be offered in a physical space. Communities that embrace knowledge sharing find that members can benefit from different experiences and perspectives that are not available in their immediate vicinity and can lead to peer education \cite{forsyth2024}. Supportive online communication has been shown to increase members' life satisfaction and overall happiness when they feel more connected to other individuals~\cite{chen2022engage, kelly2022supporting}. 

Specifically, within software development, online communities have established themselves as almost essential. Online communities, within and outside of software development and technology, foster a place of belonging and fulfillment of social needs \cite{lampe2010motivations}. In particular, small online communities provide an accessible hub for information exchange and socialization and an audience for other members, fostering discussion and providing an outlet for group identity \cite{chen2019participants,hwang2021people, teblunthuis2022no}. No single community can provide everything for its user base, but when communities are used with other communities, it opens members up to more extensive and perceivable benefits~\cite{hwang2021people, teblunthuis2022no}.

%% file: sections/03_method.tex
\section{Methodology}
This section discusses our study design, participants, and data collection and analysis. 

\subsection{Study Design}
We conducted this study with a survey and interview component, with the interview being an additional and optional route for viewers to participate in. We designed our survey questions using Qualtrics\footnote{https://www.qualtrics.com/} to elicit qualitative and quantitative data related to stream viewership. The survey begins by collecting basic demographic data on age, location, years of programming experience, level of education, and how long participants have been watching development live streams. Next, participants move into specific questions about their viewership habits of development live streams, how they found the streamers they watch, and what aspects of streams are enjoyable or beneficial. Finally, at the end of the survey, there is a voluntary opportunity to sign up and participate in the interview portion of the research. Semi-structured interview questions were designed to elicit qualitative data focused on more in-depth data on motivations to watch development live streams, interactions within the streams, and what participants feel they receive from watching streams. Each interview started with basic demographic questions, programming experience, and level of education and then move into a semi-structured question-and-answer format. We asked participants about what software and game development live streams they watch, if and how they participate in the streams, how development live streams have impacted their personal development practices, and if they are involved in the streamer's community outside of the streaming platform.

We take a reflexive thematic analysis approach to our qualitative data and analyze the open-ended survey questions and interviews separately, but discuss the resulting themes together. Through the lens of interpretivism and constructivism, we plan to take the experiences and perceptions of live stream viewers and situate them within the context of explaining and understanding developers' needs and perceived outcomes from live streamed content. Using Braun and Clarke's six-phase process, we familiarized ourselves with the data, generated initial codes using a subset of the survey and interview data, generated initial themes, reviewed our data given these initial themes, finalized and refined themes given our codes and potential sub-codes, and then summarize our findings given the produced themes \cite{braun2021one, braun2021thematic}.

Clemson's Institutional Review Board (IRB) has reviewed and granted permission to conduct this human subjects research, and participants are not offered an incentive for participation.

\subsection{Interview Design}
Semi-structured interview questions were designed to elicit qualitative data. Interviews centered around motivations to watch development live streams, interactions within the streams, and what participants felt they received from watching streams. Each interview started with basic demographic questions, programming experience, and level of education and then moved into a semi-structured question and answer format. We asked participants about what development live streams they watch, if and how they participate in the streams, how development live streams have impacted their own personal development practices, and if they are involved in the streamer's community outside of the streaming platform. 

Interview data collected builds on and uses the data from surveys to understand viewers' motivations to watch live streams, more in-depth data about the type and content of development streams they prefer, and community involvement. Using both the survey and interview data, we provide a holistic view of viewership motivations and preferences and provide a better understanding of where learning opportunities may occur within live streams for viewers. 

\subsection{Survey Design}
Survey data collected includes viewers' programming and education background, how long they have been watching software and game development live streams, how often they watch live streams, and what aspects of the streams are enjoyable to them. We also ask questions regarding community engagement and self-views on watching live streams, adapting our self-view questions from Hiranrat \emph{et al.} \cite{hiranrat2021theory}. These self-view questions help ground and provide insights into motivations and perceived benefits of watching and participating in live streamed development content. Next, we investigate online social support using the online social support scale from Nick \emph{et al.} \cite{nick2018online}. Understanding the online social support viewers potentially receive from streamers, their content, or other viewers provides additional insights into the motivations and benefits viewers may have. Example survey and interview can be found in the Appendix.

\subsection{Participant Recruitment}
To recruit participants, we posted in relevant subreddits and Discord servers where software and game developer communities are, and that allow the posting of research inquiries. Additionally, we reached out directly to streamers with known communities through email or Discord messages and worked with streamers to gain access to their communities to post our survey and engage with their viewers. The rationale behind reaching out to streamers before posting in communal spaces is out of respect for their brand and that many of these spaces, while public, are often hyper-specific and close-knit groups of individuals with shared common bond interest in the streamer and their content. As newcomers and researchers, we feel it is unethical to invade these close-knit communities without prior permission from server owners and/or the streamer just to gather data. In addition to this, we did not post directly in chat on streams for two reasons: 1) streamers often have link-sharing in chat off to prevent spam and scams, and 2) it is disingenuous and disruptive to streams to post advertisements. We believe that this self-imposed and respectful restriction has led to a smaller than expected number of participants. 

Participant inclusion criteria are any individual over the age of 18 who has watched a software or game development stream in the last 3 months or 90 days. We understand that developers may have different schedules and time commitments; therefore, we are including those individuals who may not watch or participate in a live stream as frequently. 

In total, we recorded 55 survey responses, and kept 37 responses after data cleaning and ensuring completion of essential questions. The median time to complete the survey was 11.6 minutes, and the average time to complete the survey was 41.5 minutes. Interview scheduling was available up to one week after the survey close date. In total, we scheduled 10 interviews and completed 9. The average time for interviews was 23.25 minutes.

\subsection{Participant Demographics}
The average age of survey participants was 30.5 years old, with a median age of 31; the youngest viewer was 18, and the oldest viewer was 47. The majority (92.3\%) of responses were male, with two female, one non-binary, and one participant declining to answer. Participants resided in thirteen different countries across North and South America, Europe, and India. The largest percentage of survey participants resided in the United States (46.2\%), followed by Germany (12.8\%), and the United Kingdom (7.7\%). Fourteen participants (35.9\%) held a 4-year college or university degree in computer or information science-related fields, seven (17.7\%) held a master's degree, seven (17.7\%) identified as being self-taught, seven (17.7\%) with some college or no degree, two with a 2-year degree, one with a high school/GED equivalent. Participants had an average of 8.4 years of experience programming, and the majority of participants indicated they are employed as a software or game developer (46.2\%). Nearly one-third of participants started watching software and game development streams within the last year, with about one-third starting viewership in 2018 and the longest viewer since 2012.

In total, we interviewed nine individuals who watch and or participated in software and game development live streams. The average age of interview participants was 30.7 years old, with 18 years old as the youngest participant and 46 as the oldest. Seven participants identified as male, one as non-binary, and one as female. In total, interview participants resided in four countries, with the majority (6) residing in the United States, two in Europe, and one in the United Kingdom. Two participants held held a 4-year college or university degree in computer or science, technology, engineering, or math (STEM) specialty, two held a masters degree in computer science or STEM, one held a masters in a non-STEM field, one held an associates degree in computer science, and three were computer science or STEM undergraduate students.

\subsection{Analysis}
The interview transcripts are treated as qualitative data and analyzed using reflexive thematic analysis and qualitative analysis techniques \cite{mcdonald2019reliability, braun2021thematic}. The main themes identified in the initial three interviews are then used to identify similar and additional themes in subsequent interviews. These themes are organized into broader categories and frequency of occurrence to identify potential relationships between the themes. We reviewed the high-level themes and sub-themes related to viewership, and they related to motivations for watching live streaming content, viewing patterns and preferences, streamer selection criteria, community engagement, and knowledge transfer.

Survey data consisted of both qualitative and quantitative questions. Qualitative questions were open-ended, and the team used open coding techniques to identify themes within these questions \cite{corbin2014basics}. Quantitative questions were analyzed using standard statistical methods outlined in \cite{ali2016basic}. Using the open-ended answers, we team-coded a subset of the responses to generate codes and then coded the rest of the open-ended survey questions. Finally, we see if they matched any codes from the interview analysis. 

\subsection{Limitations}
We acknowledge that several limitations and validity threats exist in this proposed work. One limitation and threat to internal validity is the potential for response bias due to how we have reached out to streamers and their communities, which may lead to socially desirable responses or pressure to participate. Additionally, this approach may lead to self-selection bias as streamers interested in the research may choose to participate and encourage their community to participate due to their influence within their community. However, approaching viewers outside their streamer's communities proves difficult and is limited by streaming platform abilities to message viewers and not be considered spam solicitation directly. Working with participants in situ in their communities can also pose variability or influence the response and behaviors of the participants. Additionally, recall bias may also play a factor in both open-ended survey and interview responses due to participants not remembering past events or experiences accurately could impact the reliability of their responses. However, due to reaching out to these communities, we anticipate that most participants gathered directly from streamer community hubs are active viewers of the streams, which we also screen participants for the last time they watched or participated in a live stream.

Another potential limitation in this research is a smaller sample size. Based on previous research experiences in software development spheres, accessing participants can be difficult with pushback from developer communities banning researchers. In our case, we are reaching out to communities centered around an individual streamer and wanted to work with streamers to access their community. This is a more respectful way to approach human-subjects research and respects communities by working through established community leaders who can vouch for the research's legitimacy and value. While potentially yielding a smaller sample, this approach aligns with ethical research practices and helps build trust between researchers and developer communities, paving the way for future studies.

%% file: sections/04_results.tex
\section{Results}
In this section, we discuss the results of viewer surveys and interviews. We outline our identified themes within the primary motivations to view live streams, viewers' perceptions of streamers and how they engage with the streamer, and viewers' perceived benefits of watching and/or participating in live streams. 

\subsection{$RQ_1$: Motivation to View Live Stream Software and Game Development}
%Survey Questions 6, 9
%Interview Questions, 7, 12

We asked survey participants to rank what motivates them to watch software and game development live streams from most important to least important motivation factors of -- learning new techniques and skills, staying updated on industry trends, entertainment and enjoyment, interacting with the streamer and the community, finding inspiration for personal projects, and other self-described motivations. Overall, most viewers reported interacting with the streamer and community as the most important primary motivator for choosing live streamed software and entertainment, while enjoyment of development content was a second-close motivating factor. Figure \ref{motivaiton} shows a full breakdown of responses.

For those most motivated by entertainment and enjoyment of a live stream, viewers often enjoyed a streamer's personality and demeanor, the amount of activity and engagement activities within the stream, appreciated having fun while gaining knowledge, and enjoyed creative problem-solving approaches. 

Interestingly, some viewers elaborated on the same interests and motivations of viewership but ranked motivational factors differently. For example, viewers who ranked entertainment and enjoyment as the most important aspect of content consumption, elaborated that they may find ``new ideas for my own project'' ranked learning lower at a level 5 or ``good community and interactivity ... motivates me to pay attention and learn,'' who ranked learning at an importance level of 4.

Next, for viewers who were most motivated by interacting with the streamer and the community at large, viewers noted they appreciated having like-minded individuals to converse with, the ability to reduce isolation from remote work through co-working, being able to give back to the community with their own knowledge and experiences, as well as engaging in respectful discussions and problem solving. 

\myquote{``I tend to work on my own for projects, so it's nice to tune into a stream on my second monitor to combat that solitude and feel like I'm working on projects. It feels like there are other devs in the room with me, just bantering as we work, which I find motivating.'' \emph{-S5}}{}

Next, for viewers most motivated to learn new techniques and skills, viewers want to see and learn from problems encountered in real time, understand how others organize knowledge and information, see different approaches to similar problems, and receive advice from others whom they perceive as more experienced.

\myquote{``As a student, it allows me to easily get advice from extremely experienced and tenured software engineers for free. ''\emph{-S22}}{}

Outside of how viewers ranked what motivated them the most, the major themes we found elaborated in the open-ended responses around motivation to view live streamed development content were - community and social connections, co-working environments, learning and skill development, entertainment with an educational value, and personal project motivation and inspiration. 

Overall, for survey participants, while many appreciate the technical learning aspects of live streamed software and game development, the social aspects and sense of community are equally, if not more important, motivators based on viewers' rankings. Many use live streams as both an educational resource and a way to feel connected with others in the broader development community. 

Interview participants were asked directly what motivates them to watch software and game development content, as well as what they enjoy about these types of streams and how they interact with streams influences which streamers they choose to watch. Like survey responses, interviewees were generally motivated by learning new concepts, professional development, finding community and entertainment, and having a productive co-working environment. 

Several interviewees mentioned they were motivated by learning new concepts and sought out software development live streams as they were learning, like I4: 

\myquote{``I was still like a Newbie programmer myself. And I used to watch, you know, big Unity, mostly because I was working my internship at Unity back. Then I would learn stuff that I saw on screen like, Oh, hey! That's that's really useful for the thing I'm doing.'' \emph{I4}}{}

Community and professional development motivations go hand in hand with finding mentorship within the community and being a mentor for the community. For those motivated by community, several feel the responsibility to give back to the community that embraced and helped them in some form or another, whether that is through contributions to the streamers' development or to the community. Interviewee 7 (I7), who has participated in a particular streamer's community for several years ``know[s] the community'' and ``feel[s] safe enough to engage'' with both other viewers and with the streamers' development as well:

\myquote{``I sort of have stuck around is that I've always had that sense of I could always help with the other side of what he's doing and the design side. So I kind of looked for that opportunity to be like. Oh, I could help out with the ui design or web design, or a graphic or something, and so have that feeling of contributing'' \emph{-I7}}{}

Still, for others, like I3, the community aspect of the streams has an appealing draw:

\myquote{``[Streams are] like a community center of people who want to hang out and chat with each other. Finding people who are all doing similar things and just having a community, it's really interesting.'' \emph{-I3}}{}

\pgfplotsset{compat=1.16}
\definecolor{Entertainment}{RGB}{185,220,15}
\definecolor{Interaction}{RGB}{233,171,100}
\definecolor{Learning}{RGB}{124,174,255}
\definecolor{Inspo}{RGB}{185,220,165}
\definecolor{Updated}{RGB}{185,20,165}
\definecolor{Other}{RGB}{184,36,33}
\newcommand{\ac}[1]{\textcolor{red}{#1}}

\begin{figure*}[t!]
\centering
\resizebox{\textwidth}{!}{
\begin{tikzpicture}
\begin{axis}[clip=true,
    xbar stacked,
    legend style={
    legend columns=6,
        at={(xticklabel cs:0.5)},yshift=-5mm,
        anchor=north,
        draw=none
    },
    ytick=data,
    xtick=data,
    axis y line*=none,
    axis x line*=bottom,
    xlabel= Percentage,
    ylabel= Motivational Factors,
    tick label style={font=\footnotesize},
    legend style={font=\footnotesize},
    label style={font=\normalsize},
    xtick={0,10,20,30,40,50,60,70,80,90,100},
    width=.9\textwidth,
    bar width=6mm,
    yticklabels={Other, Staying updated on\\ industry trends , Inspiration for \\personal projects , Learning new \\techniques and skills, Interaction with \\streamer/community, Entertainment and \\enjoyment},
    yticklabel style={font=\small, align=right},
    xmin=0,
    xmax=100,
    area legend,
    y=8mm,
    enlarge y limits={abs=0.625},
    visualization depends on=x \as \rawx,
    every node near coord/.style={align=center,text=black, font=\small},
    nodes near coords={\pgfmathprintnumber{\pgfplotspointmeta}\%}, % <-- prints % sign after y coordinate value
    xticklabel={\pgfmathprintnumber{\tick}\%},% <-- prints % sign after x tick value
    %nodes near coords align={center} % <-- horizontal alignment centered of nodes 
]
\addplot %rank 1
[Entertainment,fill=Entertainment] 
coordinates
{(0,0) (2.7,1) (5.4,2) (18.9,3) (37.8,4) (35.1,5)};
\addplot %rank 2
[Interaction,fill=Interaction] 
coordinates
{(2.7,0) (8.1,1) (21.6,2) (24.3,3) (16.2,4) (27,5)};
\addplot %rank 3
[Learning,fill=Learning] 
coordinates 
{(8.1,0) (13.5,1) (18.9,2) (24.3,3) (21.6,4) (13.5,5)};
\addplot %rank 4
[Inspo,fill=Inspo] 
coordinates 
{(0,0) (29.7,1) (32.4,2) (16.2,3) (5.4,4) (16.2,5)};
\addplot %rank 6
[Updated,fill=Updated] 
coordinates 
{(0,0) (40.5,1) (18.9,2) (16.2,3) (16.2,4) (8.1,5)};
\addplot %Rank 6
[Other,fill=Other,] 
coordinates 
{(89.2,0) (5.4,1) (2.7,2) (0,3) (2.7,4) (0,5)};

% X, 0 = Other(the category)
%X, 1, updated

\legend{Rank 1 (Most), Rank 2, Rank 3, Rank 4, Rank 5, Rank 6 (Least)}
\end{axis}  
\end{tikzpicture}
}
\caption[. ]
    {\tabular[t]{@{}l@{}} Visualization of each motivational factor to watch live streaming ranked from most to least important to viewers.\endtabular}
\label{motivaiton}
\end{figure*}
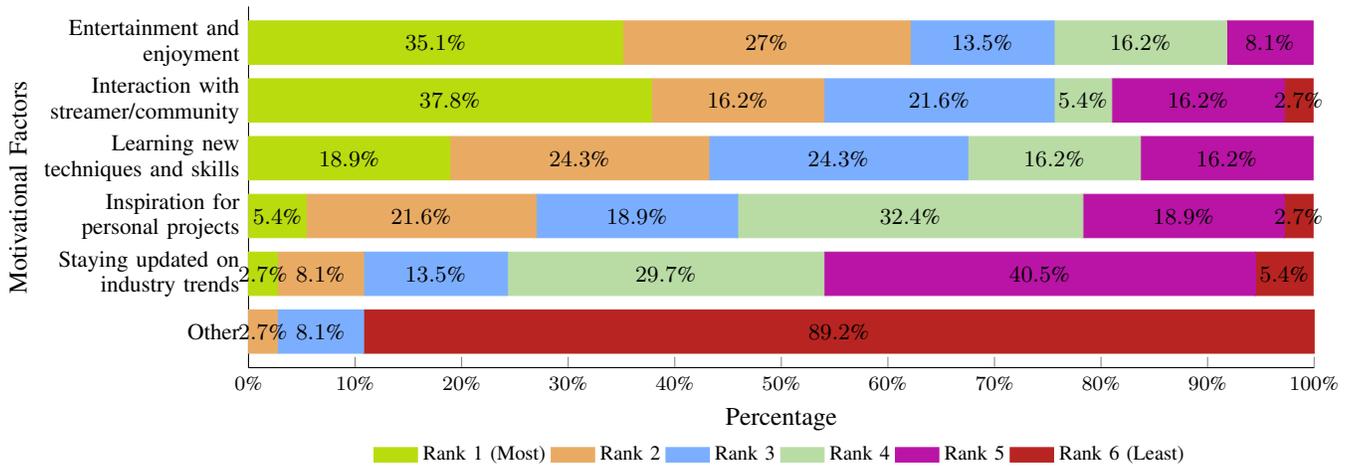

\subsection{$RQ_2$: Perceptions of Streamers and Engaging with Stream}
%Survey Qs 2, 10
%Interview 7, 8, 11, 12
To better understand viewers' perceptions of the live streamers they watch, we asked participants if the streamers they watched do work similar to themselves or how it was different. Almost half of participants, 48.7\%, indicated that they do not watch streamers who do work similar to themselves, 29.2\% indicated that streamers so work similar to themselves, and 23.1\% indicated that it varies on who they watch if the work is similar or different. For those who indicated that stream content was different from their own work, they watch game development streams despite being a full-stack, web, or otherwise not a game developer. For some participants, streams ``sometimes line up fairly well with my hobby work'' and other participants want to ``mix it up to understand new things.'' For those who watched streams with content similar to their own daily work, many were game developers who chose to watch other game developers. 

\myquote{``[Specific streamer] works in a different stack, but we both work on games in general.'' \emph{-S23}}{}

We also asked participants how they interacted in their streams to better understand how interaction preferences might reveal why and how they choose to participate in streams and the perceptions of a stream as a whole. The majority of participants (72.5\%) indicated that they are active in chat during a live stream and participated in joining a related Discord or other online community related to the stream (70\%). We identified that participants valued the ability to contribute or be a member of a community, whether that was during stream or offline on another platform. Several participants used streams as a co-working environment and learning community as they saw the value in peer-mentors, general entertainment value, and access to subject matter experts. 

Interview data sought to expand and articulate perceptions of streamers. We found three overarching themes articulated regarding perceptions of streamers to be mentorship, technical expertise, and streamer personality engagement style. 

Notably distinct from survey participants, interviewee participants perceive streamers to be mentors and that active or passive participation in streams and outside communities contributes to professional development and growth as a developer. Viewers see their streamers as role models and developer advocates and value technical knowledge and professional practices displayed by the streamers. 

\myquote{``His stream is like a mentorship type thing. This is a guy that's at the top of his field. He does stuff really well. He's got good processes. He speaks his mind and he thinks out loud so you can hear his thought processes and why he's making decisions around what he's doing.'' \emph{-I7}}{}

Multiple interviewees, however, note that stream size and mentorship opportunities may be influenced by community size and overall engagement from other viewers. With some noting how quality engagement with streamers has changed, or even depart from larger streams because the tone of streamers and visibility of viewers changed. 

\myquote{``Before [streamer] blew up I was actively engaged in [their community]. It's just impossible now, that community used to have like 5,000 people on Discord. Now, it's 1 million or something like you can't engage in it. There's communities where there's just so much happening you can't stay involved.'' \emph{-I4}}{}

\myquote{``[I] like smaller streams with fewer people talking where I can actually interact with the streamer instead of larger streamers, where it sort of feels like shouting into the void.'' \emph{-I9}}{}

Interview participants noted that they viewed streamers as someone with more technical experience than them and often had the ability to teach well, even if the streams were not necessarily educational in nature. Streamers were perceived as capable to communicate concepts effectively and, as I1 says ``showing the messy process of developing.'' Viewers have an appreciation when streamers can demonstrate real-time problem-solving that includes mistakes and the debugging process. 

Personality and engagement style can influence viewer perceptions. Many noted that a streamer's ability to maintain an engaging but comfortable atmosphere was valuable. I7 described how they make small attempts at interaction to gauge the streamer's responsiveness and personality, and this initial experience would determine whether they become an active participant or remain a lurker in a stream. I3 noted that the ability for streamers to interact with chat and go back into work seamlessly was beneficial, but acknowledged that it takes skill and not all streamers can achieve this level of interaction. 

\subsection{$RQ_3$:Perceived Benefits}
%Survey Qs 3
$RQ_3$ seeks to provide details about potential perceptible benefits of live stream viewership in both social and technical aspects. From a technical perspective, viewers believe they can benefit from a real-time learning experience watching problem-solving as issues arise, learning new techniques directly from those they perceive as subject matter experts, and receiving knowledge from other experienced developers. From a social perspective, streams reduced isolation from remote work through a co-working environment, connections with like-minded individuals, and the ability to form mentor and mentee relationships.  

In addition to the benefits identified through viewers' motivations to watch live streamed content and perceptions of engaging within a stream, we also sought to understand if watching a live stream directly benefited a viewer in their daily development practices by asking if they have used knowledge gained from a live stream directly in their work. Over half (58.\%) of participants reported that they have used knowledge or technique from a stream directly in their work, with 33.3\% stating they have not, and 7.7\% reporting that they are unsure or cannot remember. 

%Interview Qs 7, 9, 10, 13
Interview participants' benefits surrounded three major themes -- educational value, motivation, and community. The most shared beneficial aspect of streams was educational value for viewers. Viewers appreciated the opportunity and space to learn new tools, technology, and problem-solving approaches through observation. I9 mentions that it is beneficial for them to be able to see how others organize their projects or think through features:

\myquote{``[I like to] see how their project is put together, how they're making features.'' \emph{-I9}}{}

Observational learning provides value and realistic insight into development processes and professional work. Educational aspects also extend to border professional development and mentorship opportunities, as discussed above in perceptions of streamers.  

The next theme observed was the motivational productivity benefits of live stream viewership. Many use live streams as a form of co-working and virtual community spaces. Streams can provide ambient noise and a sense of accountability, even virtually. Having others present in the background while working alleviates some solitude from remote work and keeps participants focused and on task. 

\myquote{``I do tend to watch streamers for stream many hours at a time and I can just park in that stream because that's where you know the community is'' \emph{-I3}}{}

\myquote{``It keeps me on task to have someone nearby also doing work and being productive, even if that person is like on a live stream, and not actually present.'' \emph{-I9}}{}

Next, the third theme centered around community and social connection. While viewers may have initially watched for technical content, many stayed for the relationships built with streamers and other community members. These communities provided safe spaces for asking questions, networking with other developers, and engaging in both technical and social discussions. Notably, the value of this community aspect often extended beyond the live streams themselves into Discord servers and other platforms, though engagement levels varied significantly.

\myquote{``I like that interaction. Not only that's, you know, related to code, but also just in general, speaking to people who are like me'' \emph{-I9}}{}

Finally, two supporting themes emerge that coincide with each of the three major themes -- entertainment and inspiration. Viewers appreciated streamers who could balance educational content with an engaging personality and community. They also found value in seeing different approaches to projects and implementation strategies, which provides inspiration for their own work.

%% file: sections/05_discussion.tex
\section{Discussion}
In this paper, through surveys and interviews, we answered three research questions showing that viewers live streaming software and game development content 1) have multiple motivations to watch and engage with a streamer, 2) perceive streamers to be mentors and similar to themselves, and 3) share several benefits of watching streamed content. This section discusses the observed themes and phenomena and how live streaming is a potential informal learning opportunity for developers. 

\subsection{Motivation and Engagement Patterns}
%motivation notes - dual value, passive benefits even when not engaging with content, community reciprocity, professional development 
Live streaming platforms for software and game development serve a dual value as viewers seek both social connections and technical knowledge exchange from streams, and typically not one over the other. Several participants noted that their primary motivation when seeking a live stream was for educational content, but over time they transitioned to a primary entertainment or community focused viewing. This finding is consistent with prior research by Chen \emph{et al.}, who found that viewers often approach live streams without specific problems to solve or concrete learning objectives \cite{chen2021towards}. However, as we observed, viewers' evolving motivations suggest that while technical learning may initially draw viewers to development streams, the combination of education, motivation, community, and entertainment creates a uniquely valuable experience that keeps them engaged over time.

The role of community appeared as a central theme and motivator throughout, though engagement patterns varied from active to lurking, and stream size plays an important role for engagement. Smaller streams facilitate more direct interaction between viewers and streamers, enabling meaningful discussions and relationship building. Many participants also chose to join off-stream communities related to the streamers, but engaged in a selective manner using them primarily for announcements or technical help and not social interaction.

%RQ2 and perceptions of streamers 
\subsection{Viewer's Perceptions of Streamers}
Viewers' perceptions of streamers follow one of four patterns - reinforcing knowledge in a viewer's current technology stacks, learning about new technologies they may not use in their current work, gaining general programming knowledge and inspiration even if working in a different domain, and staying connected to projects they are interested in. We believe that the variety in viewers use and motivation for watching streams suggests that similarly to a streamer is not a primary factor in choosing what streamers to watch, as many deliberately seek out different content from their daily work. 

Viewer interaction preferences helped us to understand the different ways viewers chose to participate in streams, with a mix of active and passive participation for viewers and shows how viewers perceive their own role in the streaming environment and what level of engagement they are comfortable with. Viewers who participate in streamers for community integrations often see the streamer as a mentor or teacher, the stream as a place for collaboration or entertainment, and a venue for subject matter experts. These aspects indicate that streaming spaces provide a virtual co-working environment or learning community. These engagement patterns also reveal how viewers value the content they choose to consume and if they choose to learn from it through active or passive participation. Understanding these preferences helps inform viewer-streamer relationships and how viewers position themselves within the broader streaming community, which is crucial for understanding any social and educational benefits of software development live streams. 

\subsection{Benefits of Live Stream Viewership}
Many of the benefits noted by participants overlapped with motivations and reinforced other benefits. Educational, social, and entertainment aspects work together to create value for individuals. We also see viewers engaging in a stream for one benefit, like education, but stay for others like the community or entertainment. This suggests that live streams can serve multiple complimentary purposes for viewers, regardless of their original intentions. These benefits mirror findings from Faas \emph{et al}., who observed that live programming streams create a collaborative mentorship environment where knowledge flows not just from streamers to viewers but also among viewers and from viewers to streamers \cite{faas2018watch}. Haaranen also noted that programming streams during events like Ludum Dare (a game development competition) revealed rich discussions among viewers about technical topics, showing how streams serve as venues for knowledge exchange even beyond direct streamer-viewer interactions \cite{haaranen2017programming}.

Interestingly, the benefits identified through this work appear to persist regardless of the viewer's experience level - from complete beginners to experienced professionals, each finding value in different aspects of the streams. This suggests that live streaming serves as a multifaceted platform that can simultaneously support technical learning, professional development, community building, and entertainment in the software development space.

\subsection{Live Streams for Informal Education}
Live streaming provides an informal learning outlet that can complement traditional education, offering real-time exposure to professional developers, development practices, and problem-solving strategies. Live streams afford learners a ``messy'' and real look at development processes, benefiting all levels of learners.

The community aspect of streams creates a support network for learners that can be engaged with at an individual's preferred pace and comfort level. Technical and social aspects of live stream may create a sustainable learning environment where viewers can benefit from both streamer expertise and peer knowledge exchange within the community. Co-working aspects of streams can also help maintain motivation and accountability for viewers to keep learning, as streams have been shown to be beneficial for streamers themselves \cite{kokinda2023streaming}. We believe this flexibility and accessibility, combined with the ability to learn across different technology domains without formal prerequisites, makes live streaming a valuable complement to traditional software development education. We also see that the communities that surround live streams and live streamers find similar benefits and models after Denson \emph{et al}'s findings that informal learning environments provide a place for informal mentorship, camaraderie and community, and space for time management and accomplishment to be found \cite{denson2015benefits}.

This work, coupled with Haaranen's and Kokinda's previous findings, begins to build evidence that live streaming may fill important gaps in formal computing education by providing access to authentic development practices, exposing learners to real-time problem-solving approaches, and fostering communities of practice where knowledge is collaboratively constructed \cite{haaranen2017programming, kokinda2023streaming}. These platforms may be particularly valuable for democratizing access to programming education, as they require minimal equipment and preparation compared to formal course development, potentially reaching learners who might otherwise lack access to quality computing education resources.

%% file: sections/06_conclusion.tex
\section{Conclusion}
This paper presents the results of 39 surveys and nine interviews regarding software and game development live streaming motivations, habits, and perceptions. The findings indicate that viewers of live streams are primarily motivated by community interaction and educational opportunities, with many viewers valuing both social and technical knowledge gained from the respective communities they watch. Analysis of both survey and interview data shows that viewers often perceive streamers as mentors and subject matter experts, while also appreciating the authentic presentation of development processes. We outlined several key benefits of live stream viewership to be reducing isolation of remote work through a pseudo-co-working environment, opportunities for both passive and active learning, and access to developer communities. Live streaming provides a valuable informal learning platform that could complement traditional education by offering real-time exposure to professional development practices while fostering a supportive community where viewers can engage based on their own comfort levels. 

%% file: sections/07_surveyqs.tex
\section{Open-Ended Interview Questions and Survey Questions for Viewers}
\subsection{Interview Questions}
\noindent Demographic information:
\begin{itemize}
    \item Name:
    \item Age:
    \item Gender you identify as:
    \item Location:
    \item Background Education:
    \item How long have you been programming?
    \item Can you describe your work/experience background - not looking for specific company but are you a teacher, developer for a company?
    \item What languages do you typically code in?
    \item Out of the languages you mentioned, which languages would you say you are strongest in?
    \item What are your motivations for programming? / Why do you program?
\end{itemize}

\noindent General Questions
\begin{itemize}
    \item What types of software or game development live streams do you typically watch?
    \item How did you find the live streaming developers who you like to watch?
    \item What do you like about watching live streamed software development?
    \item Do you have any favorite software or game development live streamers? If so, who are they, and what do you like about their content?
    \item Have you ever used knowledge or skills gained from software and game development live streams in your own projects or work? If so, could you provide an example?
    \item When you watch live streams related to software and game development, do you typically prefer longer, more in-depth sessions or shorter, more focused content? Please explain your preference.
    \item What improvements or changes would you like to see in the live streams you watch related to software and game development?
\end{itemize}

\noindent Questions about how they engage with streamers: 
\begin{itemize}
    \item How they interact with streamers influence which streams they pick?
    \item Are you involved with a live streamers community “off-stream”, for example a Discord, Subreddit, or Patreon page?
    \begin{itemize}
        \item[]If yes:
        \item What prompted you to join the community?
        \item What are the most helpful aspects of the the community?
    \end{itemize}
    \item Do you regularly contribute to the community? If so, how?
    \item Are these contributions related to technical or programming?
    \item Are they related to social or other community interests?
    \item By being a member of this community, what do you expect to get out of your membership and participation there?
    \item What keeps you coming back to the community
\end{itemize}

\noindent Additional questions:
\begin{itemize}
    \item Do you stream?
    \item Software and game development or something else?
\end{itemize}

\subsection{Survey Questions}
\noindent Demographic questions:
\begin{itemize}
    \item Age:
    \item Gender you identify as:
    \item Location:
    \item Do you have a formal education in computer science, computer engineering, software engineering/development, information science? (Drop down)
    \item Which of the following best describes you as a developer: (Student, employed professional software or game developer, hobbyist, other)
\end{itemize}

\noindent Background on Programming:
\begin{itemize}
    \item How long have you been programming for?
    \item How long have you been watching software and game development live streams?
    \item When did you start watching software and game development live streams?
\end{itemize}

\noindent General Streaming Questions:
\begin{itemize}
    \item What types of live streams do you typically watch? 
        \begin{itemize}
            \item Software development
            \item Web development
            \item Game development
            \item Other:
        \end{itemize}
    \item Of the streamers you watch, would you consider the work they do similar to yours? (Short answer)
    \item Have you ever used a technique or solution that you have seen on a stream in your own work?  (Short answer)
    \item How often do you watch live streams related to software and game development? 
        \begin{itemize}
            \item Daily, weekly, monthly, rarely, never
        \end{itemize}
    \item What platforms do you primarily use to watch live streams(select all that apply)
    \begin{itemize}
        \item Twitch, youtube, FB gaming, Mixer, Other
    \end{itemize}
    \item  How do you engage with the community while watching live streams? (Select all that apply)
     \begin{itemize}
         \item Chatting in the live stream's chat
         \item Participating in live polls and quizzes
         \item Joining related Discord or other online communities
         \item Sharing the live stream on social media
         \item Other: 
     \end{itemize}
    \item What motivates you to watch software and game development live streams? 
        \begin{itemize}
            \item Learning new techniques and skills
            Staying updated on industry trends
            Entertainment and enjoyment
            Interacting with the streamer and the community
            Finding inspiration for personal projects
            Other:
        \end{itemize}
        \item Are there specific software development languages or game development engines you prefer to watch live streams about? If so, please specify. (Short answer)
        \item How do you discover new live streams related to software and game development? (Select all that apply)
         \begin{itemize}
             \item Recommendations from friends
             \item Social media recommendations
             \item Streaming platform recommendations
             \item Online forums and communities
             \item Search engine results
             \item Other:
         \end{itemize}
    \item What factors make a live stream enjoyable for you? (Select all that apply)
         \begin{itemize}
             \item Quality of content and information shared
             \item Interaction with the audience (e.g., Q\&A, chat engagement)
             \item Streamer's personality and presentation style
             \item Consistent schedule and reliability
             \item Length of the live stream
             \item Other:
         \end{itemize}
    \item Are there any specific topics or types of content you wish to see more of in live streams related to software and game development? (Short answer)
\end{itemize}